\documentclass[conference]{IEEEtran}
\IEEEoverridecommandlockouts
\usepackage[utf8]{inputenc}
\usepackage{cite}
\usepackage{overpic}
\usepackage{amsmath,amssymb,amsfonts}
\usepackage{bm}
\usepackage{graphicx}
\usepackage{textcomp}
\usepackage{xcolor}
\usepackage{float}
\usepackage{amsthm}
\usepackage{graphicx}
\usepackage{epstopdf}
\usepackage{amsmath,bm,bbm}
\usepackage{amsfonts}
\usepackage{amssymb}
\usepackage{color}
\usepackage{multirow}
\usepackage{multicol}
\usepackage{soul,xcolor}
\usepackage{setspace}

\usepackage{longtable}
\usepackage{mathtools}
\usepackage{subfig}
\usepackage{tabulary}
\usepackage{epsfig}
\usepackage{caption}
\usepackage{booktabs}
\usepackage{blindtext}
\usepackage{adjustbox}
\usepackage{hyperref}
\usepackage{dirtytalk}
\usepackage{comment} 

\usepackage{tikz}
\usetikzlibrary{positioning} 
\usepackage{algorithm}
\usepackage{algpseudocode}

\definecolor{mypurple}{HTML}{9933FF}
\definecolor{mygreen}{HTML}{009900}

\theoremstyle{plain}

\newcommand{\vect}[1]{\mathbf{#1}}

\def\Htran{\mbox{\tiny $\mathrm{H}$}}
\def\Ttran{\mbox{\tiny $\mathrm{T}$}}

\begin{document}
\bstctlcite{IEEEexample:BSTcontrol}
\makeatletter
\newcommand*{\rom}[1]{\expandafter\@slowromancap\romannumeral #1@}
\makeatother

\title{ Joint Quantized Precoding and Bit Allocation for Fronthaul-Constrained Cell-Free Massive MIMO  \vspace{-0.4cm}  }

\author{\IEEEauthorblockN{\"Ozlem Tu\u{g}fe Demir}
\IEEEauthorblockA{\textit{Department of Electrical and Electronics Engineering} \\
\textit{Bilkent University}\\
Ankara, Turkiye \\
E-mail: ozlemtugfedemir@bilkent.edu.tr} 
 
\thanks{This work was carried out within the scope of the project 122C149 – Intelligent End-to-End Design of Energy-Efficient and Hardware Impairments-Aware Cell-Free Massive MIMO for Beyond 5G. \"O. T. Demir was supported by the 2232-B International Fellowship for Early Stage Researchers Programme funded by the Scientific and Technological Research Council of Türkiye (TÜBİTAK).
}}
\maketitle
\begin{abstract}
We study quantization-aware precoding for the downlink of cell-free massive MIMO systems with limited-resolution fronthaul. In such systems, the centrally designed precoder must be quantized before being conveyed to distributed access points (APs), creating a strong coupling between precoder design and fronthaul compression. To capture this effect, we develop an end-to-end signal model based on Bussgang decomposition, where the quantization distortion depends explicitly on the precoder coefficients. Building on this model, we formulate a joint precoding and bit allocation problem under per-AP power constraints and a total fronthaul bit budget. We propose an efficient block coordinate descent algorithm that iteratively updates the precoder, receive scaling, and distortion levels. Numerical results demonstrate that the proposed method significantly outperforms uniform bit allocation and quantization-unaware precoding schemes, particularly in fronthaul-limited regimes.
\end{abstract}

\vspace{-2mm}
\section{Introduction}

Cell-free massive multiple-input multiple-output (MIMO) has emerged as a key technology for beyond-5G and 6G systems due to its ability to provide uniformly high spectral efficiency by jointly serving user equipments (UEs) with a large number of distributed access points (APs)~\cite{cell-free-book}. In practical deployments, the baseband processing is often split between distributed radio units and a centralized processing unit, connected via capacity-limited fronthaul links, which introduce additional quantization distortion~\cite{khorsandmanesh2023optimized,masoumi2019performance}.

While early research on quantized precoding mainly focused on low-resolution digital-to-analog converters (DACs) at the transmitter side~\cite{Jacobsson2017a,Mezghani2009-fy,Yuan2020-hz,Saxena2016-gs,Park2019-sz}, a more recent line of work has highlighted that quantization over the fronthaul—where the precoder or processed signals are compressed before being forwarded to distributed units—constitutes a fundamentally different and practically relevant bottleneck \cite{khorsandmanesh2023optimized,khorsandmanesh2026quantized}. In such architectures, the centrally designed precoder is quantized before being conveyed over the fronthaul links to the distributed APs, which creates a direct coupling between precoder design and fronthaul quantization.

Recent studies have shown that adapting the quantization resolution across links or data streams can provide substantial performance gains over uniform allocation strategies~\cite{stream-adaptive}. However, these works primarily focus on co-located MIMO systems and do not capture the structure of downlink cell-free massive MIMO, where a centrally designed precoder must be distributed across many APs under fronthaul constraints. Although the bit allocation problem has also been considered in~\cite{kim2024meta}, that work focuses on quantization of the received signal in the uplink rather than on quantized precoder design in the downlink for a cell-free massive MIMO system.

In the downlink of cell-free massive MIMO, the precoder is designed at the CPU and then quantized before being forwarded to the APs. This creates a strong coupling between precoding and fronthaul compression, which is not accounted for in conventional approaches that design the precoder assuming ideal fronthaul and apply quantization afterward. As a result, directly quantizing a conventional precoder leads to suboptimal performance under fronthaul constraints.

In this paper, we develop a quantization-aware precoding framework for the downlink of cell-free massive MIMO systems with limited-resolution fronthaul. By adopting an end-to-end signal model based on Bussgang decomposition, we explicitly characterize the impact of fronthaul quantization on the transmitted signals. Building on this model, we propose a joint optimization approach that adapts both the precoder and the per-AP per-UE quantization resolutions under a total fronthaul bit constraint.

The main contributions of this paper are summarized as follows:
\begin{itemize}
    \item We develop an end-to-end quantization-aware system model for the downlink of cell-free massive MIMO systems that captures the coupling between precoding and fronthaul quantization.
    
    \item We formulate a joint precoding and bit allocation problem under per-AP power constraints and a total fronthaul bit budget.
    
    \item We propose an efficient block coordinate descent algorithm that jointly updates the precoder, receive scaling, and quantization distortion levels.

\end{itemize}

\section{System Model and Problem Formulation}

We consider the downlink of a cell-free massive MIMO system with $L$ distributed APs, each equipped with $N$ antennas, jointly serving $K$ single-antenna UEs. The total number of transmit antennas is $M = LN$. All APs are connected to a CPU via capacity-limited fronthaul links. The CPU has access to global channel state information (CSI), which is assumed to be perfectly known unless otherwise stated, and is used to design the downlink precoder. The aggregated channel matrix is defined as
\begin{align}
\vect{H}
=
\begin{bmatrix}
\vect{H}_1 & \cdots & \vect{H}_L
\end{bmatrix}
\in \mathbb{C}^{K \times M},
\end{align}
where $\vect{H}_\ell \in \mathbb{C}^{K \times N}$ represents the channel from AP $\ell$ to all UEs.

The downlink precoder is partitioned across the APs as
\begin{align}
\vect{P}
=
\begin{bmatrix}
\vect{P}_1^{\Ttran} & \cdots & \vect{P}_L^{\Ttran}
\end{bmatrix}^{\Ttran}
\in \mathbb{C}^{M \times K},
\end{align}
where $\vect{P}_\ell \in \mathbb{C}^{N \times K}$ is the precoding matrix corresponding to AP $\ell$, and its $k$th column is denoted by $\vect{p}_{\ell k} \in \mathbb{C}^{N}$.

The transmitted data vector is $\vect{s} \in \mathbb{C}^{K}$ with normalized power
\begin{align}
\mathbb{E}\{\vect{s}\vect{s}^{\Htran}\} = \vect{I}_K.
\end{align}

\subsection{Fronthaul Quantization Model}

Due to the limited-capacity fronthaul links between the CPU and the APs, the precoder coefficients are quantized before being forwarded to the APs. We adopt a scalar Lloyd--Max quantization model and employ a Bussgang-type decomposition. The quantized precoder vector corresponding to AP $\ell$ and UE $k$ is modeled as
\begin{align}
\widehat{\vect{p}}_{\ell k}
=
(1-\eta_{\ell k}) \vect{p}_{\ell k}
+
\vect{q}_{\ell k},
\label{eq:quantized_branch_model_revised}
\end{align}
where:
 $\eta_{\ell k} \in (0,1)$ is the distortion factor determined by the quantization resolution, and $\vect{q}_{\ell k} \in \mathbb{C}^{N}$ is the quantization distortion term.

The Bussgang decomposition implies that $\vect{q}_{\ell k}$ is uncorrelated with $\vect{p}_{\ell k}$, but not statistically independent. Moreover, the distortion is generally spatially correlated across antennas. For analytical tractability, such correlations are neglected and a diagonal approximation will be adopted in the sequel.

At the AP, a normalization is applied to compensate for the attenuation factor $(1-\eta_{\ell k})$, yielding
\begin{align}
\widetilde{\vect{p}}_{\ell k}
=
\frac{\widehat{\vect{p}}_{\ell k}}{1-\eta_{\ell k}}
=
\vect{p}_{\ell k}
+
\underbrace{\frac{\vect{q}_{\ell k}}{1-\eta_{\ell k}}}_{\text{amplified distortion}}.
\label{eq:normalized_precoder_branch_revised}
\end{align}

In the precoder design stage, the distortion statistics are characterized through the factors $\{\eta_{\ell k}\}$, and the corresponding covariance is assumed to be known. For analytical tractability, it is further assumed that each AP--UE pair $(\ell,k)$ is associated with a strictly positive resolution parameter (i.e., $\eta_{\ell k} < 1$), which corresponds to allocating at least one quantization bit.

In contrast, in the final implementation of the proposed algorithm, the fronthaul bit allocation may assign zero bits to certain AP--UE pairs. In such cases, the corresponding precoder branch is not transmitted, which is equivalent to setting
\begin{align}
\vect{p}_{\ell k} = \vect{0}.
\end{align}
Hence, transmission from AP $\ell$ to UE $k$ only occurs if a positive number of bits is allocated to that branch.

Furthermore, during implementation, the precoder coefficients are quantized using scalar Lloyd--Max quantizers designed for Gaussian inputs. In particular, each precoder entry is modeled as a complex Gaussian random variable with variance $\frac{PL}{K M}$, where $P$ is the per-AP antenna power limit.

\medskip

Stacking all columns, the effective precoder becomes
\begin{align}
\widetilde{\vect{P}} = \vect{P} + \vect{E},
\end{align}
where $\vect{E} \in \mathbb{C}^{M \times K}$ collects the distortion terms.

\subsection{Quantization Distortion Covariance}

From the Bussgang-type model in \eqref{eq:quantized_branch_model_revised}--\eqref{eq:normalized_precoder_branch_revised}, the normalized distortion affecting the $n$th antenna coefficient of branch $(\ell,k)$ has variance
\begin{align}
\mathbb{E}\!\left\{
\left|
\left[
\frac{\vect{q}_{\ell k}}{1-\eta_{\ell k}}
\right]_n
\right|^2
\right\}
=
\frac{\eta_{\ell k}}{1-\eta_{\ell k}}
\left|[\vect{p}_{\ell k}]_n\right|^2,
\qquad n=1,\ldots,N.
\end{align}
This follows from the fact that, before normalization, each distortion element has variance
$\eta_{\ell k}(1-\eta_{\ell k}) |[\vect{p}_{\ell k}]_n|^2$.

Neglecting the spatial correlation of the distortion terms for analytical tractability, we adopt the following diagonal covariance approximation:
\begin{align}
\vect{R}_{Q,\ell k}
=
\frac{\eta_{\ell k}}{1-\eta_{\ell k}}
\,
\mathrm{diag}
\big(
|[\vect{p}_{\ell k}]_1|^2,\ldots,|[\vect{p}_{\ell k}]_N|^2
\big).
\label{eq:RQ_lk}
\end{align}

The aggregate distortion covariance matrix across all APs is then modeled as
\begin{align}
\vect{R}_Q
=\mathbb{E}\left\{\vect{E}\vect{E}^{\Htran}\right\}=
\mathrm{blkdiag}
\left(
\sum_{k=1}^K \vect{R}_{Q,1k},\,
\ldots,\,
\sum_{k=1}^K \vect{R}_{Q,Lk}
\right),
\label{eq:RQ_total_revised}
\end{align}
where different blocks of the centralized precoder are assumed to be statistically independent, and so are their corresponding quantization distortions.

\subsection{Received Signal and Power Constraint}

The transmitted signal is
\begin{align}
\vect{x} = \widetilde{\vect{P}} \vect{s},
\end{align}
and the received signal at the UEs is
\begin{align}
\vect{y}
=
\vect{H}\vect{P}\vect{s}
+
\vect{H}\vect{E}\vect{s}
+
\vect{n},
\label{eq:received_signal_revised}
\end{align}
where $\vect{n} \sim \mathcal{CN}(\vect{0}, \sigma^2 \vect{I}_K)$.

Each AP is subject to a transmit power constraint
\begin{align}
\|\vect{P}_\ell\|_F^2 \le P, \quad \forall \ell.
\label{eq:perAP_power_constraint_revised}
\end{align}

In the algorithmic development, we assume that the transmitted signal $\vect{x} = \widetilde{\vect{P}}\vect{s}$ is directly radiated by the APs without additional scaling, even though the quantization process may increase the instantaneous transmit power. Nevertheless, the per-AP power constraints are enforced through the design of the continuous (unquantized) precoder.

In the practical implementation, however, each AP enforces its power constraint after quantization. The effective transmit power at AP $\ell$ is given by
\begin{align}
P_\ell^{\mathrm{eff}} = \left\|\widetilde{\vect{P}}_\ell\right\|_F^2.
\end{align}
If $P_\ell^{\mathrm{eff}} > P$, the AP rescales its precoding matrix as
\begin{align}
\widetilde{\vect{P}}_\ell
\leftarrow
\sqrt{\frac{P}{P_\ell^{\mathrm{eff}}}}
\widetilde{\vect{P}}_\ell.
\label{eq:final_rescaling_revised}
\end{align}

\subsection{Objective Function}

To flexibly balance signal scaling and interference suppression, we introduce a diagonal receive-scaling matrix
\begin{align}
\vect{B} = \mathrm{diag}(\beta_1,\ldots,\beta_K),
\end{align}
where $\beta_k > 0$ is a real-valued scaling coefficient associated with UE $k$. We consider the mean-squared error (MSE) between the transmitted symbol vector $\vect{s}$ and the scaled received signal $\vect{B}\vect{y}$, i.e.,
\begin{align}
\mathcal{J}(\vect{P}, \boldsymbol{\eta}, \vect{B})
=
\mathbb{E}\left\{
\left\|
\vect{s} - \vect{B}\vect{y}
\right\|_2^2
\right\}.
\end{align}

By substituting the received signal model in \eqref{eq:received_signal_revised} and using $\mathbb{E}\{\vect{s}\vect{s}^{\Htran}\}=\vect{I}_K$, the objective function can be expressed as
\begin{align}
\mathcal{J}(\vect{P}, \boldsymbol{\eta}, \vect{B})
&=
\left\|
\vect{I}_K - \vect{B}\vect{H}\vect{P}
\right\|_F^2
\\
&\quad+
\mathrm{tr}
\left(
\vect{B}\vect{H}\vect{R}_Q\vect{H}^{\Htran}\vect{B}^{\Htran}
\right)
+
\sigma^2 \mathrm{tr}(\vect{B}\vect{B}^{\Htran}),
\end{align}
which jointly captures the signal distortion, the impact of quantization noise, and the thermal noise.

\subsection{Bit Allocation Model}

We relate the distortion factor $\eta_{\ell k}$ to the number of quantization bits $b_{\ell k}$ using the high-resolution approximation
\begin{align}
\eta_{\ell k} \approx c_q 2^{-2 b_{\ell k}},
\label{eq:eta_bits_relation_revised}
\end{align}
where $c_q > 0$ depends on the quantizer design and input distribution. This relation shows that increasing the number of bits exponentially reduces the distortion.

The total fronthaul bit budget is then expressed as
\begin{align}
\sum_{\ell=1}^L \sum_{k=1}^K
\frac{N}{2} \log_2 \left( \frac{c_q}{\eta_{\ell k}} \right)
\le B_{\mathrm{tot}},
\label{eq:bit_constraint_revised}
\end{align}
where each $(\ell,k)$ branch consumes $N b_{\ell k}$ bits.

\subsection{Problem Formulation}

The joint precoding and bit allocation problem is formulated as
\begin{subequations}\label{eq:problem_formulation}
\begin{align}
\min_{\{\vect{P}_\ell\},\,\{\eta_{\ell k}\},\,\vect{B}}
\quad
&
\mathcal{J}(\vect{P}, \boldsymbol{\eta}, \vect{B}) \label{eq:main_objective}
\\
\text{s.t.}\quad
&
\|\vect{P}_\ell\|_F^2 \le P, \quad \forall \ell,
\\
&
\sum_{\ell=1}^L \sum_{k=1}^K
\frac{N}{2} \log_2 \left( \frac{c_q}{\eta_{\ell k}} \right)
\le B_{\mathrm{tot}},
\\
&
0 < \eta_{\ell k} \le c_q, \quad \forall \ell,k.
\end{align}
\end{subequations}

\section{Block Coordinate Descent Algorithm}

Problem \eqref{eq:problem_formulation} is non-convex due to the coupling between $\vect{P}$, $\boldsymbol{\eta}$, and $\vect{B}$. We tackle it by block coordinate descent (BCD), where we iteratively update $\vect{B}$, the AP precoder blocks $\{\vect{P}_\ell\}$, and the distortion factors $\{\eta_{\ell k}\}$.

\subsection{Update of the Receive Scaling Matrix}

For fixed $\vect{P}$ and $\boldsymbol{\eta}$, define
\begin{align}
\vect{M} \triangleq \vect{H}\vect{P}.
\end{align}
Then, \eqref{eq:main_objective} becomes separable in $\{\beta_k\}$. The optimal update is
\begin{align}
\beta_k^\star
=
\frac{
\Re\{[\vect{M}]_{k,k}\}
}{
\|\vect{m}_k\|_2^2 + [\vect{H}\vect{R}_Q\vect{H}^{\Htran}]_{k,k} + \sigma^2
},
\qquad k=1,\ldots,K,
\label{eq:beta_update}
\end{align}
where $\vect{m}_k^{\Ttran}$ is the $k$th row of $\vect{M}$.\footnote{The positivity constraint $\beta_k > 0$ is relaxed in the derivation for analytical tractability. In numerical simulations, the resulting values are observed to be non-negative.}

\subsection{Update of the AP Precoder Block \texorpdfstring{$\vect{P}_\ell$}{Pl}}

For fixed $\vect{B}$ and $\boldsymbol{\eta}$, we update one AP block $\vect{P}_\ell$ while keeping the others fixed. Define
\begin{align}
\vect{R}_\ell
\triangleq
\vect{I}_K
-
\sum_{j\neq \ell}\vect{B}\vect{H}_j\vect{P}_j
\in\mathbb{C}^{K\times K},
\label{eq:Rl_def}
\end{align}
which is the residual after removing the contributions of all APs except AP~$\ell$. Moreover, let
\begin{align}
\vect{D}_\ell
\triangleq
\mathrm{diag}\!\left(
\mathrm{diag}\!\left(\vect{H}_\ell^{\Htran}\vect{B}^{\Htran}\vect{B}\vect{H}_\ell\right)
\right)
\in\mathbb{C}^{N\times N}.
\label{eq:Dl_def}
\end{align}
Using \eqref{eq:RQ_lk}, the block-subproblem for AP $\ell$ becomes
\begin{subequations}\label{eq:Pl_subproblem}
\begin{align}
\min_{\vect{P}_\ell}
\quad
&
\left\|
\vect{R}_\ell - \vect{B}\vect{H}_\ell \vect{P}_\ell
\right\|_F^2
+
\sum_{k=1}^K
\frac{\eta_{\ell k}}{1-\eta_{\ell k}}
\,
\vect{p}_{\ell k}^{\Htran}\vect{D}_\ell \vect{p}_{\ell k}
\\
\textnormal{s.t.}\quad
&
\|\vect{P}_\ell\|_F^2 \le P.
\end{align}
\end{subequations}

The Lagrangian of \eqref{eq:Pl_subproblem} is
\begin{align}
\mathcal{L}_\ell
&=
\left\|
\vect{R}_\ell - \vect{B}\vect{H}_\ell \vect{P}_\ell
\right\|_F^2
+
\sum_{k=1}^K
\frac{\eta_{\ell k}}{1-\eta_{\ell k}}
\,
\vect{p}_{\ell k}^{\Htran}\vect{D}_\ell \vect{p}_{\ell k}
\nonumber \\
&\quad+
\lambda_\ell
\left(
\|\vect{P}_\ell\|_F^2 - P
\right),
\end{align}
where $\lambda_\ell\ge 0$ is the Lagrange multiplier.

Let
\begin{align}
\vect{A}_\ell \triangleq \vect{H}_\ell^{\Htran}\vect{B}^{\Htran}\vect{B}\vect{H}_\ell.
\end{align}
Then, the columns of $\vect{P}_\ell$ decouple. For the $k$th column, the optimality condition gives
\begin{align}
\left(
\vect{A}_\ell
+
\frac{\eta_{\ell k}}{1-\eta_{\ell k}}\vect{D}_\ell
+
\lambda_\ell \vect{I}_N
\right)
\vect{p}_{\ell k}
=
\vect{H}_\ell^{\Htran}\vect{B}^{\Htran}\vect{r}_{\ell k},
\label{eq:Pl_column_update}
\end{align}
where $\vect{r}_{\ell k}$ is the $k$th column of $\vect{R}_\ell$. Hence,
\begin{align}
\vect{p}_{\ell k}^\star(\lambda_\ell)
=
\left(
\vect{A}_\ell
+
\frac{\eta_{\ell k}}{1-\eta_{\ell k}}\vect{D}_\ell
+
\lambda_\ell \vect{I}_N
\right)^{-1}
\vect{H}_\ell^{\Htran}\vect{B}^{\Htran}\vect{r}_{\ell k}.
\label{eq:Pl_closed_form_lambda}
\end{align}

If the unconstrained solution obtained with $\lambda_\ell=0$ satisfies the power constraint, then $\lambda_\ell=0$ is optimal. Otherwise, $\lambda_\ell>0$ is chosen such that
\begin{align}
\sum_{k=1}^K \left\|\vect{p}_{\ell k}^\star(\lambda_\ell)\right\|_2^2 = P.
\label{eq:Pl_bisection_equation}
\end{align}
Since the left-hand side of \eqref{eq:Pl_bisection_equation} is monotonically decreasing in $\lambda_\ell$, the optimal multiplier can be efficiently found via bisection search.

\subsection{Update of the Distortion Factors \texorpdfstring{$\eta_{\ell k}$}{eta_lk}}

For fixed $\vect{P}$ and $\vect{B}$, the terms depending on $\eta_{\ell k}$ are contained in the quantization-noise contribution. Define
\begin{align}
\psi_{\ell k}
\triangleq
\mathrm{tr}\!\left(
\vect{B}\vect{H}_\ell
\mathrm{diag}\!\big(\mathrm{diag}(\vect{p}_{\ell k}\vect{p}_{\ell k}^{\Htran})\big)
\vect{H}_\ell^{\Htran}\vect{B}^{\Htran}
\right).
\label{eq:psi_lk}
\end{align}
Then, the $\eta$-dependent part of the objective is
\begin{align}
\sum_{\ell=1}^L \sum_{k=1}^K
\psi_{\ell k}\frac{\eta_{\ell k}}{1-\eta_{\ell k}}.
\end{align}
Thus, the $\eta$-subproblem becomes
\begin{subequations}\label{eq:eta_subproblem}
\begin{align}
\min_{\{\eta_{\ell k}\}}
\quad
&
\sum_{\ell=1}^L \sum_{k=1}^K
\psi_{\ell k}\frac{\eta_{\ell k}}{1-\eta_{\ell k}}
\\
\textnormal{s.t.}\quad
&
\sum_{\ell=1}^L\sum_{k=1}^K
\frac{N}{2}\log_2\!\left(\frac{c_q}{\eta_{\ell k}}\right)
\le B_{\rm tot},
\\
&
0<\eta_{\ell k}\le c_q.
\end{align}
\end{subequations}

Introducing the Lagrange multiplier $\nu\ge 0$ for the bit-budget constraint, the KKT condition for each $(\ell,k)$ gives\footnote{Here, we implicitly assume that $c_q<1$, but during algorithmic implementation, it is allowed to take larger values than one.}
\begin{align}
\frac{\psi_{\ell k}}{(1-\eta_{\ell k})^2}
-
\frac{\nu N}{2\ln 2}\frac{1}{\eta_{\ell k}}
=0.
\label{eq:eta_kkt}
\end{align}
This can be rearranged into the quadratic equation
\begin{align}
\nu \frac{N}{2\ln 2}(1-\eta_{\ell k})^2 - \psi_{\ell k}\eta_{\ell k}=0.
\label{eq:eta_quadratic}
\end{align}
For a given $\nu$, the solution is obtained in semi-closed form by taking the valid root of \eqref{eq:eta_quadratic} and projecting it onto $(0,c_q]$. Denoting this solution by $\eta_{\ell k}^\star(\nu)$, the multiplier $\nu$ is then selected such that the bit-budget constraint is satisfied with equality. Since the total number of required bits is monotonically decreasing in $\nu$, the optimal $\nu$ can again be found by bisection search.

It is worth noting that the update of the distortion parameters $\{\eta_{\ell k}\}$ does not follow the classical structure of WMMSE-type algorithms. In particular, the mapping between $\eta_{\ell k}$ and the number of quantization bits is inherently discrete, which breaks the smooth optimization structure typically assumed in WMMSE formulations. As a result, the $\eta$-updates involve heuristic but practically meaningful operations.

Specifically, after solving a continuous relaxation of the bit-allocation problem, the obtained $\eta_{\ell k}$ values are projected onto a set of admissible distortion levels that correspond to finite-resolution quantizers. For instance, when $\eta_{\ell k}$ exceeds a certain threshold (e.g., $0.3634$, corresponding to 1-bit resolution quantization), it is clipped to $c_q$. Similarly, extremely small values are prevented to avoid unrealistically high bit allocations. This projection step reflects the fact that, in practice, only a finite number of quantization levels can be implemented, and thus the algorithm must reconcile continuous optimization with discrete hardware constraints.

Overall, the $\eta$-update can be interpreted as a quantization-aware resource allocation step that operates outside the standard WMMSE framework, introducing non-smooth and discrete effects that are essential for accurately modeling fronthaul-limited systems.

\section{Numerical Results}

In this section, we evaluate the sum spectral efficiency achieved by the proposed joint precoding and bit-allocation scheme and compare it with two benchmark methods: i) uniform bit allocation combined with the same iterative precoder design, and ii) uniform bit allocation with a conventional regularized zero-forcing (RZF) precoder followed by quantization. The average sum spectral efficiency is computed over $10$ random setups, each containing $200$ independent channel realizations.

We consider a cell-free massive MIMO downlink with $L$ distributed APs, each equipped with $N=4$ antennas, jointly serving $K=16$ single-antenna UEs in a square area of side length $500$\,m. The APs and UEs are dropped uniformly at random, and the vertical distance between each AP and UE is set to $10$\,m. The channels are generated according to an independent Rayleigh fading model with distance-dependent large-scale fading. In particular, for the link between AP $\ell$ and UE $k$, the large-scale fading coefficient is modeled as
\begin{align}
\beta_{k\ell}[\mathrm{dB}]
=
-32.4
-31.9\log_{10}(d_{k\ell})
-20\log_{10}(3.5)
\end{align}
which corresponds to a $3.5$\,GHz carrier \cite[Table 7.4.1-1]{3GPP5G}. We consider a $50$\,MHz bandwidth and a $5$\,dB noise figure. The per-AP transmit-power limit is set to $P_{\max}^{\rm AP}=2$\,W, and the high-resolution quantization constant is chosen as $c_q=\pi\sqrt{3}/2$ corresponding to Lloyd-Max quantizer with Gaussian input \cite{quantization}. The proposed algorithm is run for $15$ outer iterations. 

\begin{figure}[t!]
		\vspace{0.1cm}
	\begin{center}
		\includegraphics[trim={0cm 0cm 1cm 0.6cm},clip,width=8.1cm]{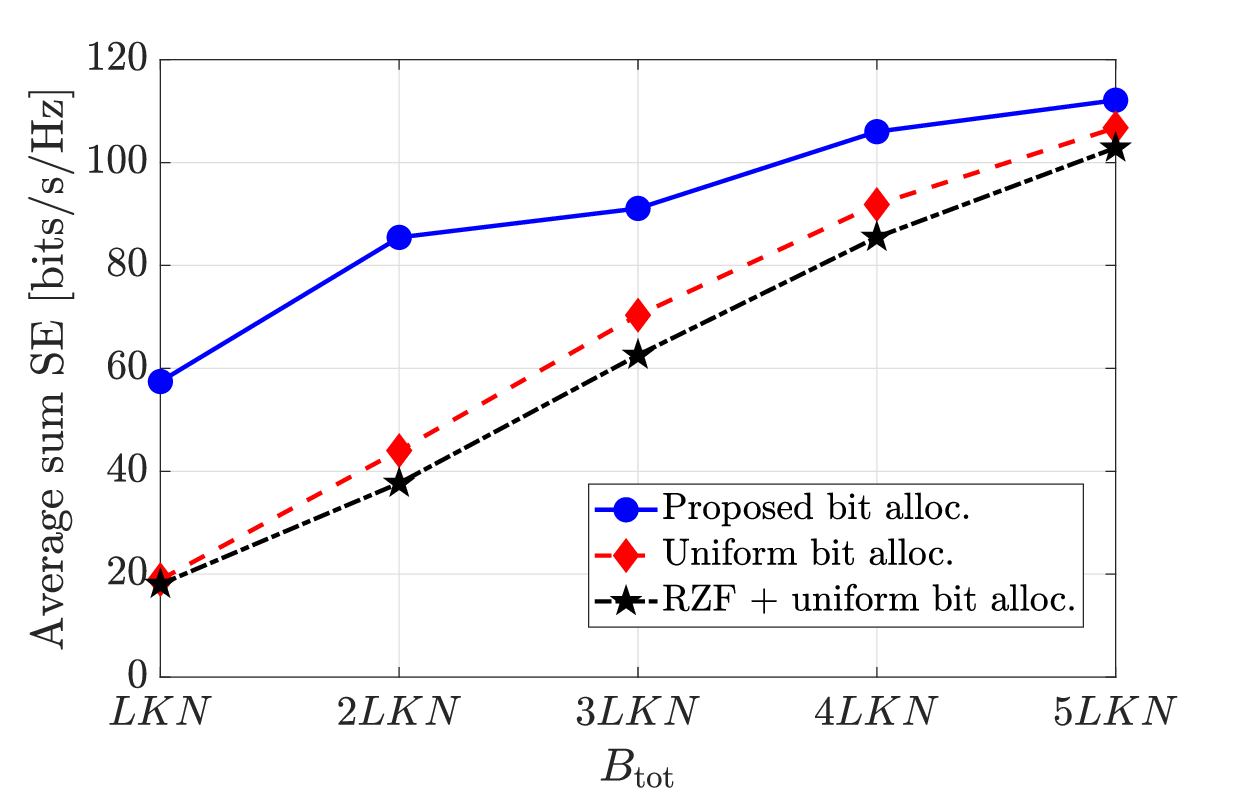}
		\caption{Average sum spectral efficiency versus the total fronthaul bit budget $B_{\rm tot}$ for $L=16$, $N=4$, and $K=16$.}
\label{fig1}
\vspace{-3mm}
	\end{center}
\end{figure}

In Fig.~\ref{fig1}, we plot the average sum spectral efficiency versus the total fronthaul bit budget $B_{\rm tot}$ for $L=16$. The results show that the proposed bit-allocation strategy consistently outperforms both benchmark schemes over the entire considered range. The gain is particularly pronounced at low and moderate fronthaul budgets, where adaptive bit allocation is most beneficial. This demonstrates that assigning the available fronthaul bits unevenly across AP--UE pairs, according to their relative importance in the precoder design, yields a substantially better use of the fronthaul resources than uniform allocation. 

Moreover, the performance gap between the uniform bit allocation combined with the same iterative precoder design and the RZF-based benchmark highlights the importance of iterative, quantization-aware precoder design. In particular, even when the same uniform bit allocation is applied, directly quantizing an RZF precoder leads to a noticeable performance loss, since the precoder itself is not adapted to the fronthaul-induced distortion. In contrast, the proposed iterative algorithm jointly updates the precoder and the distortion levels, resulting in a significantly more robust design under limited fronthaul capacity.

\begin{figure}[t!]
		\vspace{0.1cm}
	\begin{center}
		\includegraphics[trim={0cm 0cm 1cm 0.6cm},clip,width=8.1cm]{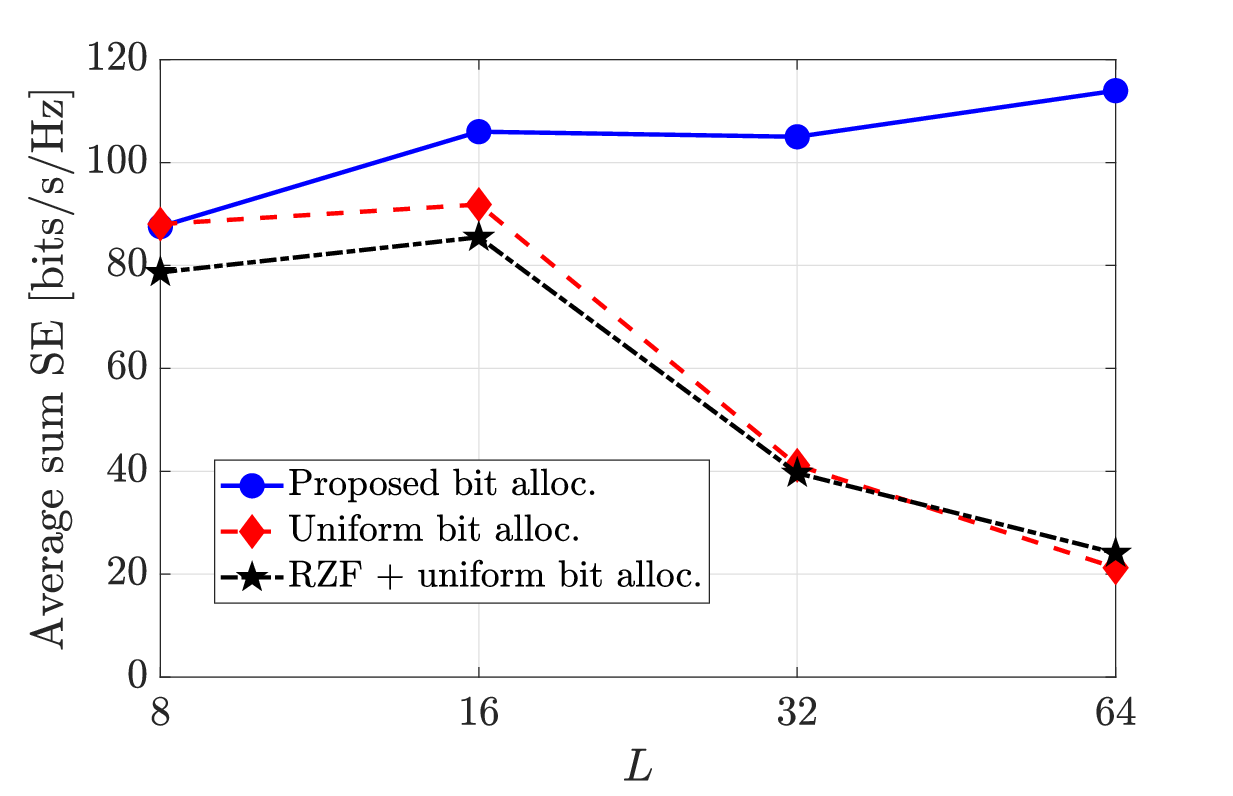}
		\caption{Average sum spectral efficiency versus the number of APs $L$ with total fronthaul budget $B_{\rm tot}=64\cdot 16 \cdot4$, where $N=4$ and $K=16$.
}
\vspace{-3mm}
\label{fig2}
	\end{center}
\end{figure}

In Fig.~\ref{fig2}, we investigate the impact of the number of APs by varying $L$ while fixing the total fronthaul bit budget to $B_{\rm tot}=64\cdot 16 \cdot4$. The proposed method maintains a high sum spectral efficiency as the network becomes denser, whereas the benchmark schemes degrade significantly for large $L$. This behavior can be explained by the fact that, under a fixed per-branch bit scaling, increasing the number of APs also increases the number of fronthaul links that must share the available quantization resources. In such cases, uniform bit allocation becomes increasingly inefficient, while the proposed method remains robust by concentrating bits on the most influential AP--UE branches. As a result, the advantage of the proposed scheme becomes even more evident in larger cell-free deployments.

\section{Conclusion}

This paper investigated quantization-aware precoding for the downlink of cell-free massive MIMO systems under fronthaul capacity constraints. By explicitly modeling the quantization of the centrally designed precoder, we revealed a strong coupling between precoding and fronthaul compression, which is not captured by conventional designs. To address this, we formulated a joint precoding and bit allocation problem and proposed an efficient block coordinate descent algorithm.

The results showed that adaptive bit allocation across AP--UE pairs provides substantial gains over uniform allocation, especially under limited fronthaul budgets. Moreover, the comparison with RZF-based benchmarks demonstrated that directly quantizing a conventional precoder leads to significant performance loss, highlighting the importance of iterative, quantization-aware design. Overall, the proposed framework provides an effective approach for improving spectral efficiency in fronthaul-constrained cell-free massive MIMO systems.

\bibliographystyle{IEEEtran}
\bibliography{IEEEabrv,refs}

\end{document}